\documentclass[twocolumn]{aastex63}
\pdfoutput=1
\usepackage{booktabs}
\usepackage{amsfonts}
\usepackage[perpage,symbol*]{footmisc}
\usepackage{color}
\usepackage{graphicx}
\usepackage{amsmath,bm}
\usepackage{gensymb}

\begin{document}

\title{Prospects for Constraining Interacting Dark Energy Models with 21 cm Intensity Mapping Experiments}

\author{Ming Zhang}
\affiliation{Department of Physics, College of Sciences, \& MOE Key Laboratory of Data Analytics and Optimization
for Smart Industry, Northeastern University, Shenyang 110819, People's Republic of China; zhangxin@mail.neu.edu.cn}
\author{Bo Wang}
\affiliation{Department of Physics, College of Sciences, \& MOE Key Laboratory of Data Analytics and Optimization
for Smart Industry, Northeastern University, Shenyang 110819, People's Republic of China; zhangxin@mail.neu.edu.cn}
\author{Peng-Ju Wu}
\affiliation{Department of Physics, College of Sciences, \& MOE Key Laboratory of Data Analytics and Optimization
for Smart Industry, Northeastern University, Shenyang 110819, People's Republic of China; zhangxin@mail.neu.edu.cn}
\author{Jing-Zhao Qi}
\affiliation{Department of Physics, College of Sciences, \& MOE Key Laboratory of Data Analytics and Optimization
for Smart Industry, Northeastern University, Shenyang 110819, People's Republic of China; zhangxin@mail.neu.edu.cn}

\author{Yidong Xu}
\affiliation{National Astronomical Observatories, Chinese Academy of Sciences, Beijing 100101, People's Republic of China; xuyd@nao.cas.cn}

\author{Jing-Fei Zhang}
\affiliation{Department of Physics, College of Sciences, \& MOE Key Laboratory of Data Analytics and Optimization
for Smart Industry, Northeastern University, Shenyang 110819, People's Republic of China; zhangxin@mail.neu.edu.cn}

\author{Xin Zhang}
\affiliation{Department of Physics, College of Sciences, \& MOE Key Laboratory of Data Analytics and Optimization
for Smart Industry, Northeastern University, Shenyang 110819, People's Republic of China; zhangxin@mail.neu.edu.cn}

\begin{abstract}
We forecast constraints on cosmological parameters in the interacting dark energy models using the mock data generated for neutral hydrogen intensity mapping (IM) experiments. In this work, we only consider the interacting dark energy models with energy transfer rate $Q=\beta H\rho_{\rm c}$, and take BINGO, FAST, SKA1-MID, and Tianlai as typical examples of the 21 cm IM experiments. We find that the Tianlai cylinder array will play an important role in constraining the interacting dark energy model. Assuming perfect foreground removal and calibration, and using the Tianlai-alone data, we obtain $\sigma(H_0)=0.19$ km s$^{-1}$ Mpc$^{-1}$, $\sigma(\Omega_{\rm m})=0.0033$ and $\sigma(\sigma_8)=0.0033$ in the I$\Lambda$CDM model, which are much better than the results of Planck+optical BAO (i.e. optical galaxy surveys). However, the Tianlai-alone data cannot provide a very tight constraint on the coupling parameter $\beta$ compared with Planck+optical BAO, while the Planck+Tianlai data can give a rather tight constraint of $\sigma(\beta)=0.00023$ due to the parameter degeneracies being well broken by the data combination. In the I$w$CDM model, we obtain $\sigma(\beta)=0.00079$ and $\sigma(w)=0.013$ from Planck+Tianlai. In addition, we also make a detailed comparison among BINGO, FAST, SKA1-MID, and Tianlai in constraining the interacting dark energy models. We show that future 21 cm IM experiments will provide a useful tool for exploring the nature of dark energy and play a significant role in measuring the coupling between dark energy and dark matter.

\end{abstract}



\section{Introduction}\label{sec1}

According to our understanding of contemporary cosmology, dark energy is responsible for the accelerating expansion of the universe \citep{Riess:1998cb}. However, the nature of dark energy still remains a deep mystery. To explore dark energy in depth and also test the possible deviation from general relativity, it is necessary to precisely measure the late-time expansion history of the universe. Among several ways of exploring the cosmic expansion history, the baryon acoustic oscillations (BAOs; \citealt{Blake:2003rh,Seo:2003pu}) have been proven to be a very useful tool to measure cosmological distances and the Hubble expansion rate and explore the nature of dark energy.
Because the acoustic waves are frozen in the eras after recombination, the BAO peak wavelength, as a cosmological standard ruler, allows accurate measurement of the expansion history.

In addition to the oscillatory features on the power spectrum of the cosmic microwave background (CMB), the BAO imprint signals as a preferred clustering scale in the large-scale structure of the later universe.
A traditional way of mapping out the large-scale structure of the universe is to use galaxies as tracers of the underlying dark matter distribution, and, by measuring the galaxy distribution with galaxy surveys, one can reconstruct the three-dimensional matter density field. However, a more efficient way to map out the matter density field over a wide range of redshifts and scales is to use neutral hydrogen (H{\tt I}) as the tracer of dark matter. The H{\tt I} pervades space from the epoch of recombination all through to the present day, and with its hyperfine-structure emission line at a rest-frame wavelength of 21 cm, we can realize the tomography of the universe over a vast cosmological volume.
In the post-reionization epochs, the bulk of H{\tt I} resides in dense gas clouds in galaxies.
Detecting a sufficiently large number of galaxies with H{\tt I} 21 cm emission could make it possible to provide a useful tool for cosmological research, but actually it is unnecessary to perform a galaxy survey for the study of large-scale structure. We can instead measure the total \rm H{\tt I} intensity over large angular scales without needing to resolve individual galaxies. Similar to the CMB map, this intensity mapping (IM) methodology makes it possible to efficiently survey large volumes with modern radio telescopes \citep{Battye:2004re,Battye:2012tg,McQuinn:2005hk,Loeb:2008hg,Pritchard:2008da,Wyithe:2007gz,Mao:2008ug,Wyithe:2007rq,Bagla:2009jy,Seo:2009fq,Lidz:2011dx,Ansari:2011bv,Bull:2014rha,Xu:2014bya,Braun:2015zta,Bull:2015lja,Bull:2015esa,Pourtsidou:2016dzn,Yohana:2019ahg,Tramonte:2019nuo,Tramonte:2020csa,Xu:2020uws,Zhang:2019dyq,Zhang:2019ipd}.

Using the \rm H{\tt I} IM method, \citet{Chang:2010jp} reported the measurements of the cross-correlation function between the \rm H{\tt I} map observed with the Green Bank Telescope (GBT) and the galaxy map observed with the DEEP2 optical redshift survey. The cross-power spectrum between the \rm H{\tt I} and optical galaxy survey was also detected with the GBT 21 cm IM survey and the WiggleZ Dark Energy Survey \citep{Masui:2012zc}. Recently, \citet{Anderson:2017ert} reported the results from 21 cm IM acquired from the Parkes radio telescope and cross-correlated with galaxy maps from the 2dF galaxy survey. So far, there are many current and future \rm H{\tt I} IM experiments comprised of wide-field and high-sensitivity radio telescopes or interferometers.
Here we consider several typical examples, namely, Baryon acoustic oscillations from Integrated Neutral Gas Observations (BINGO; \citealt{Battye:2012tg,Dickinson:2014wda,Wuensche:2019cdv,Wuensche:2019znm}), the Five-hundred-meter Aperture Spherical radio Telescope (FAST; \citealt{Nan:2011um,Smoot:2014oia,Bigot-Sazy:2015tot,Li:2012ub,Yu:2017hqz,Hu:2019okh}), the Square Kilometre Array (SKA; \citealt{Bull:2015nra,Santos:2015gra,Braun:2015zta,Braun:2019gdo}), and the Tianlai cylinder array \citep{2011SSPMA..41.1358C,2012IJMPS..12..256C,Xu:2014bya,Wu:2016vzu,Li:2020tianlaicylinder,Wu:2020jwm}.
This work aims to forecast how future \rm H{\tt I} IM experiments, including BINGO, FAST, the SKA Phase I mid-frequency array (SKA1-MID), and Tianlai, can constrain the interacting dark energy (IDE) models.

The IDE model originates from a long-standing conjecture that there might be some coupling between dark energy and cold dark matter (CDM; for a recent review, see \citealt{Wang:2016lxa}). The model with an interaction between vacuum energy (for convenience, dark energy with $w=-1$ is called the vacuum energy in this work and denoted by $\Lambda$) and CDM is usually called the I$\Lambda$CDM model. Besides, we also wish to consider a more general IDE model in which the dark energy equation of state (EoS) parameter $w$ is a constant, usually called the I$w$CDM model. In the I$w$CDM scenario, the energy conservation equations for dark energy and CDM satisfy
\begin{eqnarray}
\label{rhodedot} &&\dot{\rho}_{\rm de} = -3 H (1+w) \rho_{\rm de} + Q, \\
\label{rhocdot} &&\dot{\rho}_{\rm c} = -3 H \rho_{\rm c} - Q,
\end{eqnarray}
where $Q$ is the energy transfer rate; $\rho_{\rm de}$ and $\rho_{\rm c}$ denote the energy densities of dark energy and CDM, respectively; $H=\dot{a}/a$ represents the Hubble parameter; $a$ is the scale factor of the universe; and a dot represents the derivative with respect to the cosmic time $t$. Here the case of $w=-1$ corresponds to the I$\Lambda$CDM model. Many specific forms for $Q$ have been constructed and discussed in previous works \citep{Amendola:1999er,Zhang:2005rg,Zhang:2005rj,Zhang:2004gc,Barrow:2006hia,Zhang:2007uh,He:2008tn,He:2009mz,Valiviita:2009nu,Boehmer:2008av,Kim:2015iba,Xia:2009zzb,Wei:2010cs,Li:2011ga,Clemson:2011an,Li:2013bya,Li:2014eha,Li:2014cee,Zhang:2017ize,Yin:2015pqa,Wang:2014oga,Geng:2015ara,Valiviita:2015dfa,DiValentino:2019jae,Guo:2017hea,Guo:2018gyo,Guo:2018ans,Yang:2018euj,Feng:2017usu,Feng:2019mym,Li:2018ydj,Li:2019ajo,DiValentino:2019ffd,Zhao:2018fjj,Li:2020gtk,2019PhRvD..99b3528X,Mukhopadhyay:2020bml,2020JCAP...05..038L}. In this paper, we employ a phenomenological form of $Q = \beta H \rho_{\rm c}$, where $\beta$ denotes a dimensionless coupling parameter. From Eqs.~(\ref{rhodedot}) and (\ref{rhocdot}), it can be seen that $\beta>0$ indicates that CDM decays into dark energy,  $\beta<0$ indicates that dark energy decays into CDM, and $\beta=0$ means that there is no interaction between dark energy and CDM.

In this work, we study what role the 21 cm IM experiments would play in constraining cosmological parameters in the I$\Lambda$CDM and I$w$CDM models. Combining with the Planck CMB data \citep{Aghanim:2018eyx}, we wish to forecast how these 21 cm IM experiments will improve constraints on cosmological parameters. We also make a comparison with optical galaxy surveys. Unless otherwise stated, we employ the spatially flat $\Lambda$CDM model with parameters fixed by fitting to the Planck 2018 data \citep{Aghanim:2018eyx} as a fiducial model to generate mock data.

The constraining power of 21 cm IM experiments for the IDE models was first studied by \citet{Xu:2017rfo}. However, our work differs from the previous work in the following aspects. (i) \citet{Xu:2017rfo} considered a model with $Q = 3H(\xi_1 \rho_{\rm c} + \xi_2 \rho_{\rm de})$ with $\xi_1$ and $\xi_2$ being the dimensionless coupling coefficients, and in order to avoid the curvature perturbation divergence in early times, they excluded a part of the parameter space. In our work, we consider a different IDE model, and we do not sacrifice any part of the parameter space. To avoid the perturbation divergence, we employ an extended parameterized post-Friedmann (ePPF) framework developed by some of the authors of this work (see \citealt{Li:2014eha,Li:2014cee}). (ii) They used the angular power spectrum, while in this work we use the power spectrum $P(k)$ (actually, the BAO and redshift space distortion (RSD) information extracted from the power spectrum). (iii) They considered BINGO, FAST, and SKA, while here we also consider the Tianlai cylinder array as an additional facility, for which we will highlight a promising capability in the following sections.

This paper is organized as follows. In Section \ref{sec2}, we give a detailed description of the methodology. In Section \ref{sec2.1}, we introduce the signal power spectrum and noise power spectrum of the 21 cm IM experiments and construct the Fisher matrix. We further give a detailed description of the experimental configurations in Section \ref{sec2.2} and describe the methods and data employed in this paper in Section \ref{sec2.3}. In Section \ref{sec3}, we present forecast constraints on cosmological parameters and provide some relevant discussions. Finally, we give our conclusions in Section \ref{sec4}.

\section{Methodology}\label{sec2}

\subsection{21 cm IM}\label{sec2.1}

The mean \rm H{\tt I} brightness temperature is given by (the detailed derivation can be found in \citealt{Battye:2012tg})
\begin{eqnarray}
\overline{T}_b (z) = 180 \Omega_{\rm H{\tt I}}(z) h \frac{(1+z)^2}{H(z)/H_0} \, {\rm mK}, \label{Tb}
\end{eqnarray}
where $\Omega_{\rm H{\tt I}}(z)$ is the fractional density of \rm H{\tt I}, $H(z)$ is the Hubble parameter as a function of redshift $z$, $H_0\equiv100 h~\rm km~s^{-1}~Mpc^{-1}$ is its value today and $h$ is the dimensionless Hubble constant.
The fractional H{\tt I} density can be written as
\begin{eqnarray}
\Omega_{\rm H{\tt I}}(z) \equiv (1+z)^{-3} \rho_{\rm H{\tt I}}(z) / \rho_{c,0}, \label{Omega_HI}
\end{eqnarray}
where $\rho_{c,0}$ is the critical density today, and $\rho_{\rm H{\tt I}}(z)$ is the proper {\rm H{\tt I}} density at redshift $z$, given by
\begin{eqnarray}
\rho_{\rm H{\tt I}}(z) = \int_{M_{\rm min}}^{M_{\rm max}} dM \frac{dn}{dM} M_{\rm H{\tt I}}(M, z). \label{rho_HI}
\end{eqnarray}
Here $M$ is the dark matter halo mass, $dn/dM$ is the proper halo mass function, and $M_{\rm H{\tt I}}(M, z)$ is the {\rm H{\tt I}} mass in a halo of mass $M$ at redshift $z$. Note that the $(1+z)^{-3}$ term appears in Equation (\ref{Omega_HI}) because $dn/dM$ is the halo mass function in proper volume units. Once $M_{\rm H{\tt I}}(M, z)$ is specified, we can then obtain $\Omega_{\rm H{\tt I}}(z)$. The detailed calculations can be found in \citet{Bull:2014rha}, and the resulting $\Omega_{\rm H{\tt I}}(z)$ is shown in their Figure 20.

Considering the effect of RSDs \citep{Kaiser:1987qv}, the signal $\rm H{\tt I}$ power spectrum can be written as \citep{Seo:2003pu}
\begin{equation}
\begin{split}
P^{\rm S}(k_{ f}, \mu_{ f}, z) = & \overline{T}^2_b \frac{D_{\rm A}^2(z)_{ f} H(z)}{D_{\rm A}^2(z) H(z)_{ f}} b^2_{\rm H{\tt I}}(z) F_{\rm RSD}(k_{ f}, \mu_{ f}) P(k,z) , \label{Pk}
\end{split}
\end{equation}
where the subscript $f$ denotes the quantities calculated in the fiducial cosmology, $D_{\rm A}(z)$ is the angular diameter distance, and $b_{\rm H{\tt I}}(z)$ is the {\rm H}{\tt I} bias, calculated by
\begin{eqnarray}
b_{\rm H{\tt I}}(z) = \rho_{\rm H{\tt I}}^{-1}(z) \int_{M_{\rm min}}^{M_{\rm max}} dM \frac{dn}{dM} M_{\rm H{\tt I}}(M, z) b(M, z) , \label{bias_HI}
\end{eqnarray}
where $b(M, z)$ is the halo bias. The detailed calculations can also be found in \citet{Xu:2014bya} and \citet{Bull:2014rha}.

The RSD effect is given by $F_{\rm RSD}(k_{ f}, \mu_{ f})$ in Equation (\ref{Pk}),
\begin{equation}
\begin{split}
F_{\rm RSD}(k_{ f}, \mu_{ f}) = [1+\beta_{\rm H{\tt I}}(z) \mu^2]^2 e^{-k^2 \mu^2 \sigma_{\rm NL}^2} , \label{RSD}
\end{split}
\end{equation}
where the first term in the square bracket corresponds to the Kaiser formula, the exponential term accounts for the ``fingers of God'' effect, $\mu=\hat{k}\cdot\hat{z}$, $\beta_{\rm H{\tt I}}(z)$ is the RSD parameter equal to $f/b_{\rm H{\tt I}}(z)$ ($f\equiv d {\rm ln}D / d {\rm ln}a$ is the linear growth rate, with $a$ being the scale factor) in linear theory, and $\sigma_{\rm NL}$ is the nonlinear dispersion scale. For the fiducial model, we choose a value of $\sigma_{\rm NL} = 7~{\rm Mpc}$ \citep{Li:2007rpa}.

The matter power spectrum $P(k,z)=D^2(z)P(k,z=0)$, with $D(z)$ being the growth factor and $P(k,z=0)$ being the matter power spectrum at $z=0$ that can be generated by \texttt{CAMB} \citep{Lewis:1999bs}.

The noise power spectrum models the instrumental and sky noises for a given experiment. The survey noise properties have been described in detail in \citet{Battye:2012tg} and \citet{Bull:2014rha}. Here we summarize them for completeness. The frequency resolution of IM surveys performs well, so we ignore the instrument response function in the radial direction and only consider the response due to the finite angular resolution,
\begin{eqnarray}
W^2(k)={\rm exp}\left[-k_\perp^2r^2(z)\left(\frac{\theta_{\rm B}}{\sqrt{8{\rm ln}2}}\right)^2\right], \label{Wk}
\end{eqnarray}
where $k_\perp$ is the transverse wavevector, $r(z)$ is the comoving radial distance at redshift $z$, and $\theta_{\rm B}$ is the full width at half-maximum of the beam of an individual dish.

Considering a redshift bin between $z_1$ and $z_2$, the survey volume can be written as
\begin{eqnarray}
V_{\rm sur}=\Omega_{\rm tot} \int_{z_1}^{z_2} dz \frac{dV}{dz d\Omega} =  \Omega_{\rm tot} \int_{z_1}^{z_2} dz \frac{c r^2(z)}{H(z)}, \label{Vsur}
\end{eqnarray}
where $\Omega_{\rm tot}=S_{\rm area}$ is the solid angle of the survey area. The pixel volume $V_{\rm pix}$ is also calculated
with a similar formula, with $\Omega_{\rm tot}$ substituted by $\Omega_{\rm pix} \approx \theta^2_{\rm B}$.

For an experiment using single-dish mode, the pixel noise can be written as
\begin{eqnarray}
\sigma_{\rm pix}=\frac{T_{\rm sys}}{\sqrt{\Delta \nu \, t_{\rm tot}(\theta_{\rm B}^2/S_{\rm area})}}\frac{\lambda^2}{A_e \theta_{\rm B}^2}\frac{1}{\sqrt{N_{\rm dish} N_{\rm beam}}}, \label{Sigmapix}
\end{eqnarray}
and for an interferometer, the pixel noise can be written as
\begin{eqnarray}
\sigma_{\rm pix}=\frac{T_{\rm sys}}{\sqrt{\Delta \nu \, t_{\rm tot}({\rm FoV}/S_{\rm area})}}\frac{\lambda^2}{A_e \sqrt{\rm FoV}}\frac{1}{\sqrt{n(k_{\perp}) N_{\rm beam}}} ,\nonumber\\
\label{Sigmapix}
\end{eqnarray}
where $T_{\rm sys}$ is the system temperature, $N_{\rm dish}$ is the number of dishes, $N_{\rm beam}$ is the number of beams, $t_{\rm tot}$ is the total observing time, and $A_e$ is the effective collecting area of each element. For a dish reflector, $A_e = \eta \pi (D_{\rm dish}/2)^2$ and $\theta_{\rm B}\approx \lambda/D_{\rm dish}$, where $D_{\rm dish}$ is the diameter of the dish, and $\eta$ is an efficiency factor for which we adopt 0.7 in this work. For a cylindrical reflector, $A_e = \eta l_{\rm cyl} w_{\rm cyl}/N_{\rm feed}$ and ${\rm FoV}\approx 90^{\circ}\times \lambda/w_{\rm cyl}$, where $w_{\rm cyl}$ and $l_{\rm cyl}$ are width and length of the cylinder, respectively, and $N_{\rm feed}$ is the number of feeds per cylinder. Unlike a single dish, we need to calculate the baseline density $n(k_{\perp})$ for the interferometer. In this work, we consider all of the baselines of the Tianlai cylinder array, as detailed in \citet{Xu:2014bya}.

For BINGO, FAST, and Tianlai, the system temperature is given by
\begin{eqnarray}
T_{\rm sys} = T_{\rm rec} + T_{\rm gal} + T_{\rm CMB},
\end{eqnarray}
where $T_{\rm rec}$ is the receiver temperature for each of these experiments with the values given in Table \ref{Table1}, $T_{\rm gal} \approx 25~\rm{K} (408~\rm{MHz}/ \nu )^{2.75}$ is the contribution from the Milky Way for a given frequency $\nu$, and $T_{\rm CMB}\approx2.73~\rm K$ is the CMB temperature. For the SKA1-MID array, the system temperature is calculated by
\begin{eqnarray}
T_{\rm sys} = T_{\rm rec} + T_{\rm spl} + T_{\rm gal} + T_{\rm CMB},
\end{eqnarray}
where $T_{\rm spl}\approx3$~K is the contribution from spillover. The receiver temperature $T_{\rm rec}$ for SKA1-MID is assumed to be \citep{Bacon:2018dui}
\begin{eqnarray}
T_{\rm rec} = 15~\rm{K} + 30~\rm{K} \left(\frac{\nu}{\rm GHz}-0.75 \right)^2. \label{SKA1}
\end{eqnarray}

Finally, the noise power spectrum is then given by
\begin{eqnarray}
P^{\rm N}(k) = \sigma^2_{\rm pix}V_{\rm pix}W^{-2}(k). \label{Pnoise}
\end{eqnarray}

The Fisher matrix for a set of parameters $\{p\}$ is given by \citet{Tegmark:1997rp},
\begin{eqnarray}
F_{ij}=\frac{1}{8\pi^2}\int^{1}_{-1} d\mu \int^{k_{\rm max}}_{k_{\rm min}} k^2dk \; \frac{\partial {\rm ln}P^{\rm S}}{\partial p_i}\frac{ \partial {\rm ln}P^{\rm S}}{\partial p_j} V_{\rm eff}, \label{Fisher1}
\end{eqnarray}
where we define the ``effective volume" as \citep{Bull:2014rha,Pourtsidou:2016dzn}
\begin{eqnarray}
V_{\rm eff} = V_{\rm sur} \left(\frac{P^{\rm S}}{P^{\rm S}+P^{\rm N}}\right)^2.
\end{eqnarray}

Next, we have assumed that the bias $b_{\rm H{\tt I}}$ depends only on the redshift $z$. This assumption is appropriate only for large scales, so we impose a nonlinear cutoff at $k_{\rm max} \simeq 0.14 (1+z)^{2/3} \, {\rm Mpc}^{-1}$ \citep{Smith:2002dz}. The largest scale the survey can probe corresponds to a wavevector $k_{\rm min} \simeq 2\pi/V_{\rm sur}^{1/3}$ \citep{Smith:2002dz}. In this work, we choose the parameter set $\{p\}$ as $\{D_{\rm A}(z),H(z),[f\sigma_8](z),[b_{\rm H{\tt I}}\sigma_8](z), \sigma_{\rm NL}\}$. When inverting the Fisher matrix, we can get the covariance matrix that gives us the forecast constraint on the chosen parameter set. Note that we only use the forecast cosmological observables $D_A(z)$, $H(z)$, and $[f\sigma_8](z)$ to constrain the cosmological parameters.

\subsection{Experimental configurations}\label{sec2.2}

In this paper, we focus on the Tianlai, BINGO, FAST, and SKA1-MID experiments. These experiments are potentially suitable for the \rm H{\tt I} IM survey in the post-reionization epochs of the universe. In this subsection, we give a brief description of these experiments.

\begin{table}
\caption{Experimental configurations for Tianlai, BINGO, FAST, and SKA1-MID.}
\label{Table1}
\centering
\begin{tabular}{l|cccc}
\bottomrule[1pt]
                               & Tianlai   & BINGO     & FAST     & SKA1-MID \\
\hline
$z_{\rm min}$                  & 0         & 0.13      & 0        & 0.35 \\
$z_{\rm max}$                  & 2.55      & 0.48      & 0.35     & 3    \\
$N_{\rm dish}$                 & --         & 1         & 1        & 197  \\
$N_{\rm beam}$                 & 1         & 50        & 19       & 1    \\
$D_{\rm dish}~[\rm m]$         & --         & 40        & 300      & 15  \\
$S_{\rm area}~[\rm deg^2]$     & 10,000     & 3000      & 20,000    & 20,000  \\
$t_{\rm tot}~[\rm hr]$        & 10,000     & 10,000     & 10,000    & 10,000  \\
$T_{\rm rec}~[\rm K]$          & 50        & 50        & 20       & Eq.~(\ref{SKA1})  \\
\bottomrule[1pt]
\end{tabular}
\end{table}

$\textsl{Tianlai.}$ The Tianlai project\footnote{\url{http://tianlai.bao.ac.cn}} is an \rm H{\tt I} IM experiment aimed at measuring the dark energy EoS by detecting the BAO features in the large-scale structure power spectrum. The full-scale Tianlai cylinder array will consist of eight adjacent cylinders to be built in northwest China, with each cylinder 15 m wide and 120 m long with 256 dual polarization feeds \citep{2011SSPMA..41.1358C,2012IJMPS..12..256C,Xu:2014bya}. Note that, currently, there is a Tianlai pathfinder array commissioning, which uses a much smaller-scale cylinder array, but in this work, we will only discuss the full-scale Tianlai cylinder array that is to be built in the future.

$\textsl{BINGO.}$ The BINGO experiment\footnote{\url{http://bingotelescope.org}} is a project to build a special-purpose radio telescope to map redshifted \rm H{\tt I} emission in the redshift range of $z=0.13-0.48$. The design of BINGO is a dual-mirror compact antenna telescope with a 40 m primary mirror and an offset focus, which is proposed to have a receiver array containing 50--60 feed horns, with a 90 m focal length. It will be built in {eastern Brazil} \citep{Battye:2012tg,Dickinson:2014wda}.

$\textsl{FAST.}$ The FAST\footnote{\url{https://fast.bao.ac.cn}} is a multibeam single-dish telescope built in Guizhou province of southwest China. The aperture diameter is 500 m with an effective illuminating diameter of 300 m. It uses an active surface that adjusts shape to create parabolas in different directions. It will be capable of covering the sky within a $40\degree$ angle from the zenith. Nineteen beams are designed in one receiver array, which will greatly increase the survey speed~\citep{Nan:2011um,Smoot:2014oia}.

\begin{figure}[]
\includegraphics[scale=0.5]{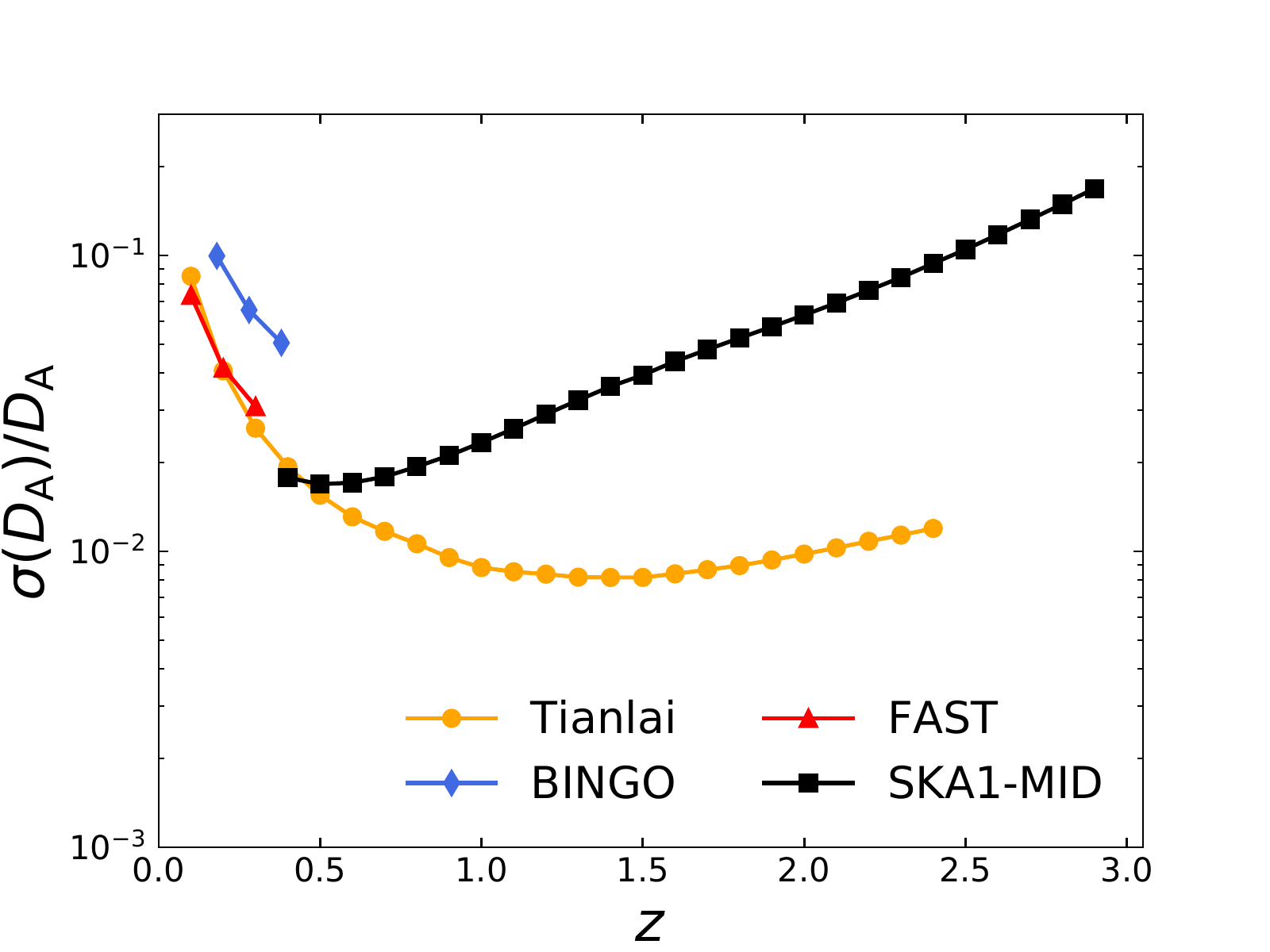}
\includegraphics[scale=0.5]{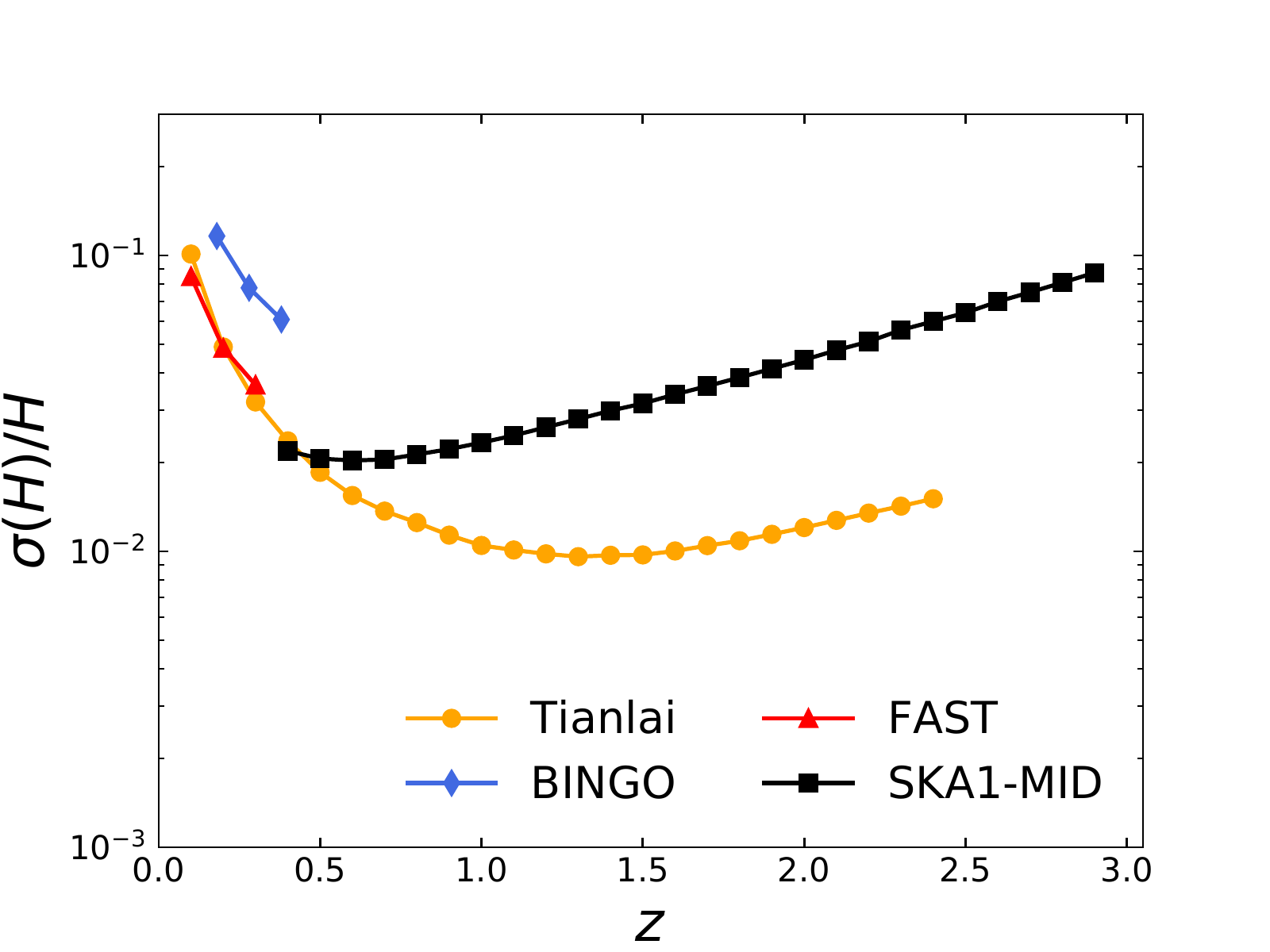}
\includegraphics[scale=0.5]{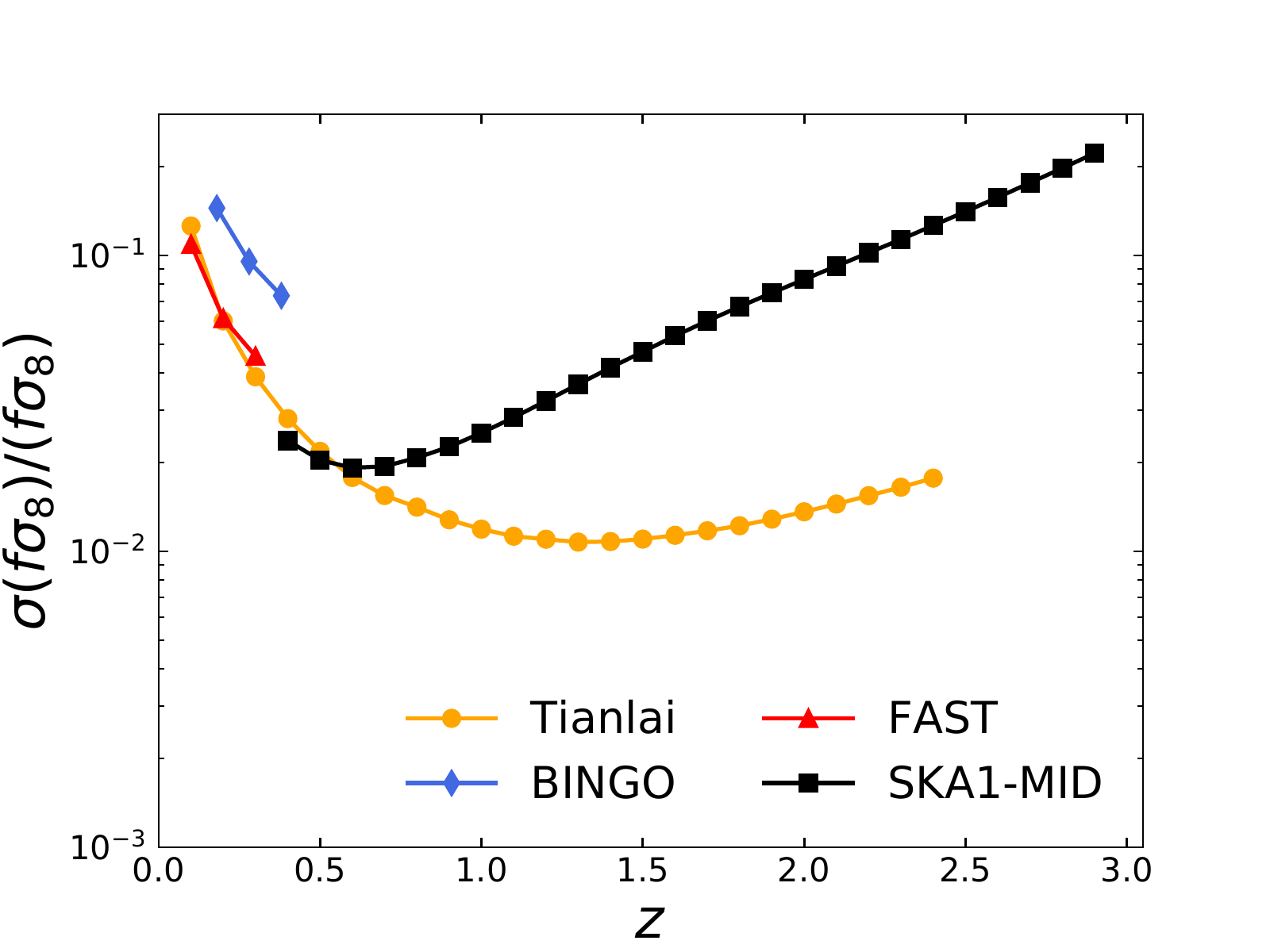}
\caption{Forecast fractional errors on $D_{\rm A}(z)$, $H(z)$, and $[f\sigma_{8}](z)$, as a function of redshift.}
\label{sigma}
\end{figure}

$\textsl{SKA1-MID.}$ The SKA project,\footnote{\url{https://www.skatelescope.org}} currently under construction, plans two stages of development. In this paper, we consider the SKA1-MID array, based in the Northern Cape, South Africa. The SKA1-MID has 133 15 m SKA dishes and 64 13.5 m MeerKAT dishes.
The SKA1-MID will perform an \rm H{\tt I} IM survey over a broad range of frequencies and a large fraction of the sky \citep{Braun:2015zta,Bull:2015nra,Santos:2015gra,Bacon:2018dui}. Here we consider only the  Wide Band 1 Survey of the SKA1-MID and use only its single-dish (autocorrelation) mode. In addition, in this work, for simplicity, we consider SKA1-MID as an array with 197 15 m dishes.

The full instrumental parameters used for these experiments are listed in Table \ref{Table1}.

\subsection{Data and method}\label{sec2.3}

A method for forecasting cosmological constraints for \rm H{\tt I} IM surveys has been presented in \citet{Bull:2014rha} and \citet{Witzemann:2017lhi}. We will follow the prescription given in \citet{Bull:2014rha} and \citet{Witzemann:2017lhi} to perform the forecast for the 21 cm IM experiments.
By performing measurements of the full anisotropic power spectrum, we obtain constraints on the angular diameter distance $D_{\rm A}(z)$, the Hubble parameter $H(z)$, and RSD observable $[f\sigma_{8}](z)$, which are considered to be independent in each redshift bin. We obtain covariance matrices for $\{D_{\rm A}(z_j), H(z_j), [f\sigma_{8}](z_j); j=1...N \}$ in a series of $N$ redshift bins $\{z_j\}$ by inverting the Fisher matrix. We perform the Fisher matrix calculations by considering the aforementioned parameters. The marginalized constraints on these parameters for these surveys are shown in Figure \ref{sigma}.

These covariance matrices, plus the fiducial cosmology, generate the mock data of these 21 cm IM experiments. Then we use these mock data to constrain the cosmological parameters by performing a Markov Chain Monte Carlo (MCMC) analysis.
The cosmological parameters that we sample include $\beta$, $H_0$, $\Omega_{\rm b}$, $\Omega_{\rm c}$, $\sigma_8$, and $w$. We consider flat priors for these parameters with ranges of $\beta\in[-1, 1]$, $H_0\in[30, 100]~\rm {km~s^{-1}~Mpc^{-1}}$, $\Omega_{\rm b}\in[0, 1]$, $\Omega_{\rm c}\in[0, 1]$, $\sigma_8\in[0, 2]$, and $w\in[-3, 1]$.
In the MCMC analysis, we also employ the CMB angular power spectra data of Planck 2018 TT,TE,EE+lowE \citep{Aghanim:2018eyx}, and the BAO measurements from galaxy redshift surveys, including SDSS-MGS \citep{Ross:2014qpa}, 6dFGS \citep{Beutler:2011hx}, and BOSS DR12 \citep{Alam:2016hwk}.

In this paper, we employ the ePPF framework to calculate the cosmological perturbations in the IDE scenario \citep{Li:2014eha,Li:2014cee}. This is because we need to avoid the perturbation divergence problem in the IDE cosmology.

It is well known that, in the IDE scenario, when calculating the cosmological perturbations, it is found that for most cases, the curvature perturbation on superhorizon scales at early times is divergent, which is a catastrophe for the IDE cosmology. The underlying reason for this problem is that we actually do not know how to consider the perturbations of dark energy. In the traditional linear perturbation theory, for calculating the perturbations of dark energy, we need to define a rest-frame sound speed for dark energy fluid $c_s^2=\delta p/\delta \rho$ (with the gauge $v|_{\rm rf}=B|_{\rm rf}=0$) to relate the dark energy density and pressure perturbations. This leads to that in a general gauge, $\delta p$ has two parts, adiabatic and nonadiabatic, and the interaction term appearing in the nonadiabatic part will occasionally lead the nonadiabatic modes to be unstable.

\begin{table}
\caption{The 1$\sigma$ errors on the parameters in the I$\Lambda$CDM model from the different data combinations. Here, $H_0$ is in units of $\rm {km~s^{-1}~Mpc^{-1}}$.}
\label{Table2}
\centering
\setlength{\tabcolsep}{1mm}{
\begin{tabular}{l|cccc}
\hline
                    & $\beta$ & $H_0$ & $\Omega_{\rm m}$ & $\sigma_8$ \\
\hline
Planck              & 0.00240 & 1.80                             & 0.0255           & 0.0150     \\
Planck+BAO          & 0.00120 & 0.69                             & 0.0087           & 0.0110     \\
Tianlai             & 0.00715 & 0.19                             & 0.0033           & 0.0033     \\
Planck+Tianlai      & 0.00023 & 0.16                             & 0.0020           & 0.0018     \\
Planck+BINGO        & 0.00170 & 1.30                             & 0.0180           & 0.0110     \\
Planck+FAST         & 0.00140 & 1.10                             & 0.0150           & 0.0089     \\
Planck+SKA1-MID     & 0.00060 & 0.39                             & 0.0054           & 0.0035     \\
\hline
\end{tabular}}
\end{table}

\begin{table}
\caption{The 1$\sigma$ errors on the parameters in the I$w$CDM model from the different data combinations.}
\label{Table3}
\centering
\begin{tabular}{l|cc}
\hline
                    & $\beta$ & $w$   \\
\hline
Planck              & 0.00175 & 0.315 \\
Planck+BAO          & 0.00150 & 0.074 \\
Tianlai             & 0.01380 & 0.016 \\
Planck+Tianlai      & 0.00079 & 0.013 \\
Planck+BINGO        & 0.00160 & 0.048 \\
Planck+FAST         & 0.00160 & 0.036 \\
Planck+SKA1-MID     & 0.00120 & 0.021 \\
\hline
\end{tabular}
\end{table}

In order to solve this problem, \citet{Li:2014eha,Li:2014cee} extended the original PPF framework \citep{Fang:2008sn} to include the IDE scenario and used this method to avoid the divergence of cosmological perturbations in the IDE models. The ePPF method does not consider the dark energy pressure perturbation; it only describes dark energy perturbations based on some basic facts of dark energy. On large scales, far beyond the horizon, the relationship between the velocities of dark energy and other components can be established through a parameterization provided by a function $f_\zeta(a)$. On small scales, deep inside the horizon, dark energy is smooth enough so that it can be viewed as a pure background. We thus can use the Poisson equation to describe this limit. In order to make these two limits compatible, we introduce a dynamical function $\Gamma$ by which we can find an equation to describe the cases on all scales. In the equation of motion of $\Gamma$, a parameter $c_\Gamma$ is introduced, giving a transition scale in terms of the Hubble scale under which dark energy is smooth enough. In this equation, there is no perturbation variable of dark energy, and when the evolution of $\Gamma$ is derived, we can directly obtain the density and velocity perturbations of dark energy. Hence, this method avoids using the pressure perturbation of dark energy defined by the sound speed. Here we only give a very brief description of the ePPF method, and we refer the reader to \citet{Li:2014eha,Li:2014cee} for more details.

We employ the \texttt{CosmoMC} package \citep{Lewis:2002ah} to perform the MCMC calculations and insert the ePPF code as a part of it to treat the cosmological perturbations in the IDE models.

\section{Results}\label{sec3}

In this section, we will present the forecast results showing relative constraining capabilities for the I$\Lambda$CDM and I$w$CDM models by combining each of the 21 cm IM experiments with Planck. Tables \ref{Table2} and \ref{Table3} list the 1$\sigma$ errors for the marginalized parameter constraints for the I$\Lambda$CDM and I$w$CDM models, respectively. In Section \ref{sec3.1}, we will show what role the 21 cm IM experiments, taking the example of Tianlai, could play in constraining cosmological parameters and compare with the BAO measurements from galaxy redshift surveys. In Section \ref{sec3.2}, we will compare the capabilities of constraining cosmological parameters for the different 21 cm IM experiments.

\begin{figure*}
\center
\includegraphics[scale=0.6]{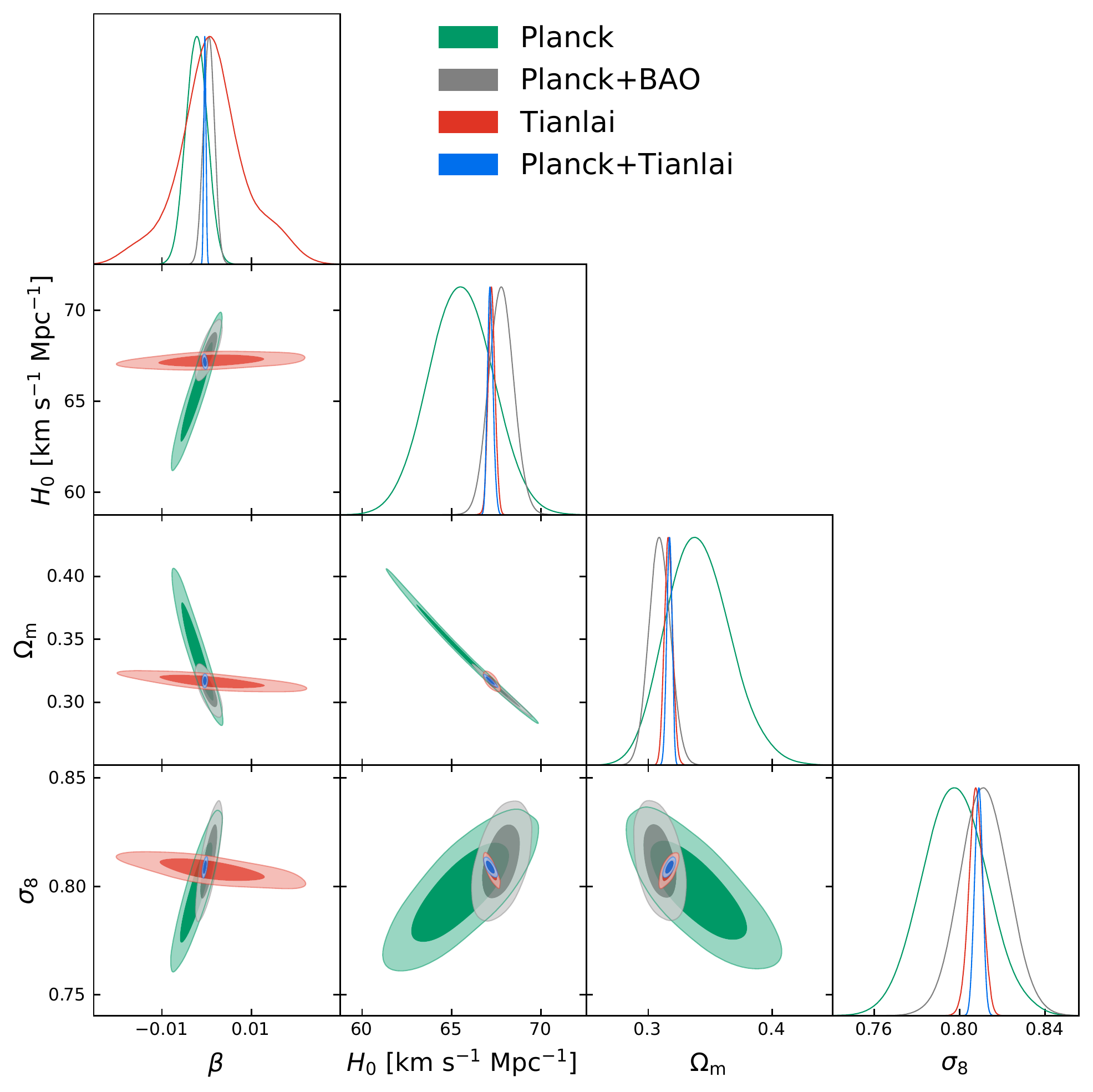}
\caption{Constraints on cosmological parameters from Planck, Planck+BAO, Tianlai, and Planck+Tianlai in the I$\Lambda$CDM model.}
\label{ILCDM1}
\end{figure*}

\subsection{Constraints on cosmological parameters from the Tianlai cylinder array}\label{sec3.1}

The 1$\sigma$ and 2$\sigma$ posterior distribution contours are shown in Figure \ref{ILCDM1} for Planck, Planck+BAO, Tianlai, and Planck+Tianlai in the I$\Lambda$CDM model. We find that the future full-scale Tianlai experiment can give very tight constraints on $H_0$, $\Omega_{\rm m}$, and $\sigma_8$. With the Tianlai data alone, we obtain $\sigma(H_0)=0.19$ km s$^{-1}$ Mpc$^{-1}$, $\sigma(\Omega_{\rm m})=0.0033$, and $\sigma(\sigma_8)=0.0033$, which are even much better than the results of $\sigma(H_0)=0.69$ km s$^{-1}$ Mpc$^{-1}$, $\sigma(\Omega_{\rm m})=0.0087$, and $\sigma(\sigma_8)=0.0110$ from the Planck+optical BAO. Comparing with the Planck data, the data combination Planck+Tianlai can improve the constraint accuracies of $H_0$, $\Omega_{\rm m}$, and $\sigma_8$ by $(1.80-0.16)/1.80=91.1\%$, $(0.0255-0.0020)/0.0255=92.2\%$, and $(0.0150-0.0018)/0.0150=88.0\%$, respectively.

Actually, the forecast analysis for some 21 cm IM experiments (including Tianlai) was made in \citet{Bull:2014rha}. In Table 4 of \citet{Bull:2014rha}, the results are shown for the $w_0$$w_a$CDM model with spatial curvature. It can be seen that our constraint results (for the I$\Lambda$CDM model) are stronger than theirs, although the models considered are different. The reasons lie in the following facts. (i) \citet{Bull:2014rha} considered the foreground residuals, but we considered a perfect removal of foreground in this work. (ii) They ignored some baselines (those shorter than 15 m), but we considered all of the baselines, especially for those short baselines formed by the large number of close-by feeds on the Tianlai cylinder array, which add a lot of sensitivity for low-redshift BAO signals. (ii) We employed the experimental configuration for Tianlai based on \citet{Xu:2014bya}, which is different from \citet{Bull:2014rha}. For example, the redshift range and survey area we use are $z=0$--2.55 and $S_{\rm area}=10,000$ deg$^2$, while in \citet{Bull:2014rha}, they adopted $z=0.49$--1.58 and $S_{\rm area}=25,000$ deg$^2$. In addition, we considered the combination with Planck 2018 data, while they used only a Planck prior from the Planck 2013 result. Also, \citet{Bull:2014rha} considered two more cosmological parameters in their forecast than ours.
These differences lead to the stronger constraints we have than theirs.

For the coupling parameter $\beta$, we find that the Tianlai-alone data can only give a relatively weak constraint, $\sigma(\beta)= 0.0072$; as a comparison, the Planck-alone data give a result of $\sigma(\beta)= 0.0024$.
In fact, for the IDE model with $Q=\beta H\rho_c$, the CMB data usually could provide a tight constraint on the coupling parameter $\beta$. This is because in the early universe, both $H$ and $\rho_c$ take rather high values, and the energy transfer rate $Q$ can take a moderate value even if $\beta$ is very small. Thus, the CMB data as an early-universe probe can give a relatively tight constraint on $\beta$. This is why the Planck CMB data can offer a much tighter constraint on $\beta$, compared with the case of the Tianlai data, but for the other cosmological parameters, the Tianlai data can give much better constraints.
Since the degeneracy directions of $\beta$ and other parameters for Planck and Tianlai are rather different, we can eventually obtain a rather tight constraint on $\beta$, i.e., $\sigma(\beta)=0.00023$, from the Planck+Tianlai data combination, which is much better than the result of $\sigma(\beta)=0.00120$ from the Planck+BAO data combination.

\begin{figure}
\center
\includegraphics[scale=0.4]{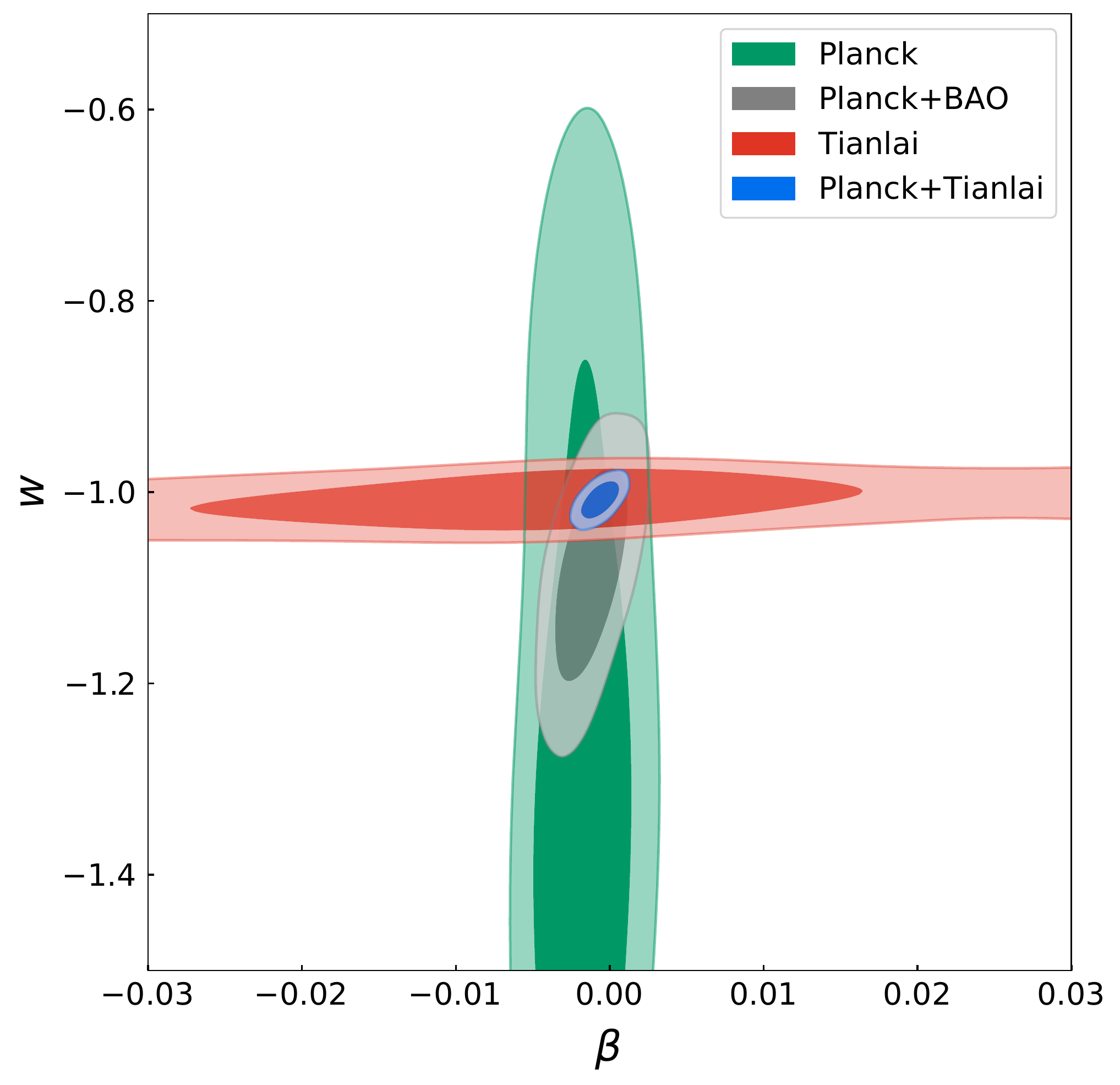}
\caption{Constraints on $\beta$ and $w$ from Planck, Planck+BAO, Tianlai, and Planck+Tianlai in the I$w$CDM model.}
\label{IwCDM1}
\end{figure}

\begin{figure*}
\center
\includegraphics[scale=0.6]{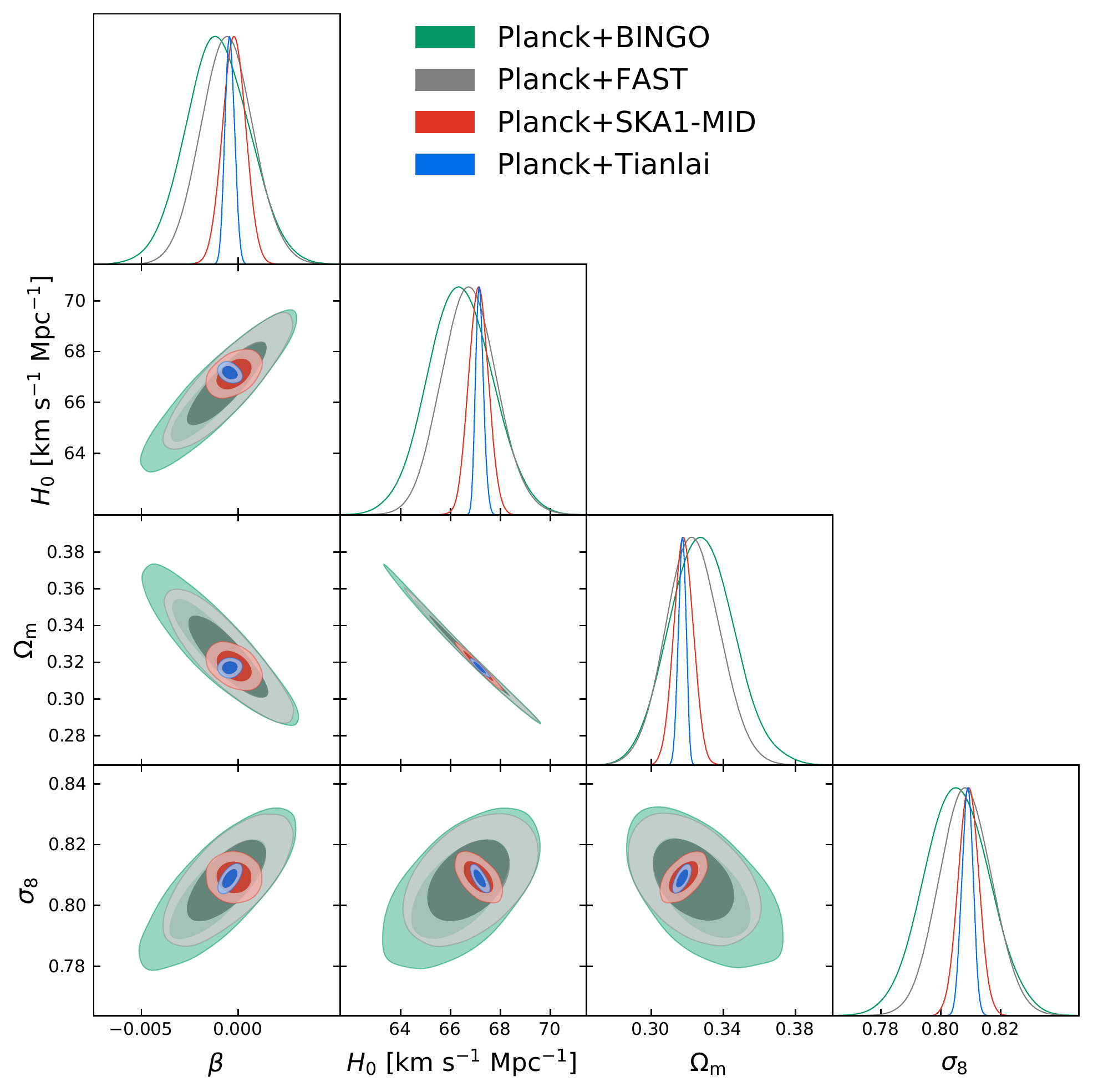}
\caption{Constraints on cosmological parameters from Planck+BINGO, Planck+FAST, Planck+SKA1-MID, and Planck+Tianlai in the I$\Lambda$CDM model.}
\label{ILCDM2}
\end{figure*}

In Figure \ref{IwCDM1}, we show the constraint results of the I$w$CDM model. Here we only show the posterior distribution contours in the $\beta$--$w$ plane that we are most interested in.
We can see that using only the Planck CMB data cannot give a good constraint on the dark energy EoS parameter $w$, i.e., $\sigma(w)\approx 0.3$. It is necessary to use the late-universe measurements to break the parameter degeneracies inherent in the CMB. As a contrast, the Tianlai-alone data can provide a rather tight constraint on $w$, giving a result of $\sigma(w)=0.016$.

However, as discussed above, the CMB data can tightly constrain the coupling parameter $\beta$ in the I$w$CDM model with the interaction term $Q=\beta H\rho_c$. So, we can see that the Planck-alone data give $\sigma(\beta)=0.00175$, which is much better than the result given by the Tianlai-alone data, $\sigma(\beta)=0.01380$.

Since the CMB data can tightly constrain $\beta$ and the Tianlai data can tightly constrain $w$, their degeneracy directions of them are entirely different. It is known that Planck alone, and even Planck+BAO, can only provide a loose or moderate constraint on the I$w$CDM model, as shown in Figure \ref{IwCDM1}. Nevertheless, since the cosmological parameter degeneracies can be broken by the Tianlai data, the parameter constraints are greatly improved by adding the Tianlai data in the fit. We obtain $\sigma(\beta)=0.00079$ and $\sigma(w)=0.013$ from the Panck+Tianlai data combination, and we find that the constraints on $\beta$ and $w$ are improved by $(0.00175-0.00079)/0.00175=54.9\%$ and $(0.315-0.013)/0.315=95.9\%$, respectively, by adding the Tianlai data. Comparing with the constraint results of $\sigma(\beta)=0.00150$ and $\sigma(w)=0.074$ from Planck+BAO, we can see that the future 21 cm IM experiments will exhibit a powerful capability of constraining cosmological parameters.

\subsection{Comparison with constraints from different 21~cm IM experiments}\label{sec3.2}

In this subsection, we will discuss the ability to constrain cosmological parameters for the different 21 cm IM experiments.

Figure \ref{ILCDM2} visualizes the constraint results for the I$\Lambda$CDM model from each of these experiments, including BINGO, FAST, SKA1-MID, and Tianlai, combined with Planck. We can clearly see from Figure \ref{ILCDM2} that the constraining capabilities of the two arrays, i.e. Tianlai and SKA1-MID, are much better than those of the single dishes, FAST and BINGO. Comparing Tianlai and SKA1-MID, Tianlai is evidently better, and comparing FAST and BINGO, FAST is much better. As shown in Figure \ref{sigma}, FAST and BINGO can only observe in low redshifts and can only cover narrow redshift ranges, i.e., $0<z<0.35$ for FAST and $0.13<z<0.48$ for BINGO. Although the redshift range coverages are similar, comparing FAST and BINGO for the relative measurement errors on $D_{\rm A}(z)$, $H(z)$, and $[f\sigma_{8}](z)$, we find that FAST is obviously better than BINGO, mainly due to the much larger survey area and larger aperture size. Comparing Tianlai and SKA1-MID, we find that both of them can cover a wide redshift range, i.e., $0<z<2.55$ for Tianlai and $0.35<z<3$ for SKA1-MID (Wide Band 1 Survey). However, for the redshift range of $z>0.5$, the relative errors on $D_{\rm A}(z)$, $H(z)$, and $[f\sigma_{8}](z)$ of Tianlai are much smaller than those of SKA1-MID. This highlights the advantage of a compact interferometer array with a large number of receivers and explains why Tianlai's capability of constraining cosmological parameters is better than that of SKA1-MID.

Concretely, for constraining the coupling parameter $\beta$, we obtain $\sigma(\beta)=0.00170$, 0.00140, 0.00060, and 0.00023, from Planck+BINGO, Planck+FAST, Planck+SKA1-MID, and Planck+Tianlai, respectively. This shows that, for constraining $\beta$ in the I$\Lambda$CDM model, the capability of Tianlai is the best, in the sense of combining with Planck, much better than those of BINGO, FAST, and SKA1-MID.
Comparing with the results of the Planck-alone data, the Planck+BINGO, Planck+FAST, Planck+SKA1-MID, and Planck+Tianlai data can improve the constraints on $\beta$ by $29.2\%$, $41.7\%$, $75.0\%$, and $90.4\%$, respectively.

We then show the results for the constraints on the other cosmological parameters in the I$\Lambda$CDM model.
We obtain $\sigma(H_0)=1.30$ km s$^{-1}$ Mpc$^{-1}$, $\sigma(\Omega_{\rm m})=0.0180$, and $\sigma(\sigma_8)=0.0110$ from Planck+BINGO and $\sigma(H_0)=1.10$ km s$^{-1}$ Mpc$^{-1}$, $\sigma(\Omega_{\rm m})=0.0150$, and $\sigma(\sigma_8)=0.0089$ from Planck+FAST. We can see that, for constraining $H_0$, $\Omega_{\rm m}$, and $\sigma_8$, FAST performs better than BINGO. But we also notice that neither FAST nor BINGO is as powerful as Tianlai and SKA1-MID.
We obtain $\sigma(H_0)=0.39$ km s$^{-1}$ Mpc$^{-1}$, $\sigma(\Omega_{\rm m})=0.0054$, and $\sigma(\sigma_8)=0.0035$ from Planck+SKA1-MID and $\sigma(H_0)=0.16$ km s$^{-1}$ Mpc$^{-1}$, $\sigma(\Omega_{\rm m})=0.0020$, and $\sigma(\sigma_8)=0.0018$ from Planck+Tianlai. We can clearly see that in the future, the full-scale Tianlai cylinder array experiment will play a significant role in precisely measuring cosmological parameters.

Here we note that, in this work, we only discuss the Wide Band 1 Survey of the SKA1-MID. Actually, SKA1 has two other surveys targeting cosmology, i.e., the Medium-Deep Band 2 Survey (with SKA1-MID) and the Deep SKA-LOW Survey (with SKA1-LOW). The Medium-Deep Band 2 Survey covers the redshift range of $0<z<0.5$ (with $S_{\rm area}=5000 ~{\rm deg}^2$ and $t_{\rm total}=10,000$ hr), and the Deep SKA-LOW Survey covers the redshift range of $3<z<6$ (with $S_{\rm area}=100~ {\rm deg}^2$ and $t_{\rm total}=5000$ hr). It is of great interest to use the combination of the three surveys of SKA1 in measuring the expansion history of the post-reionization epoch of the universe to explore various cosmological issues.
We will leave this work for the future.

\begin{figure}
\center
\includegraphics[scale=0.4]{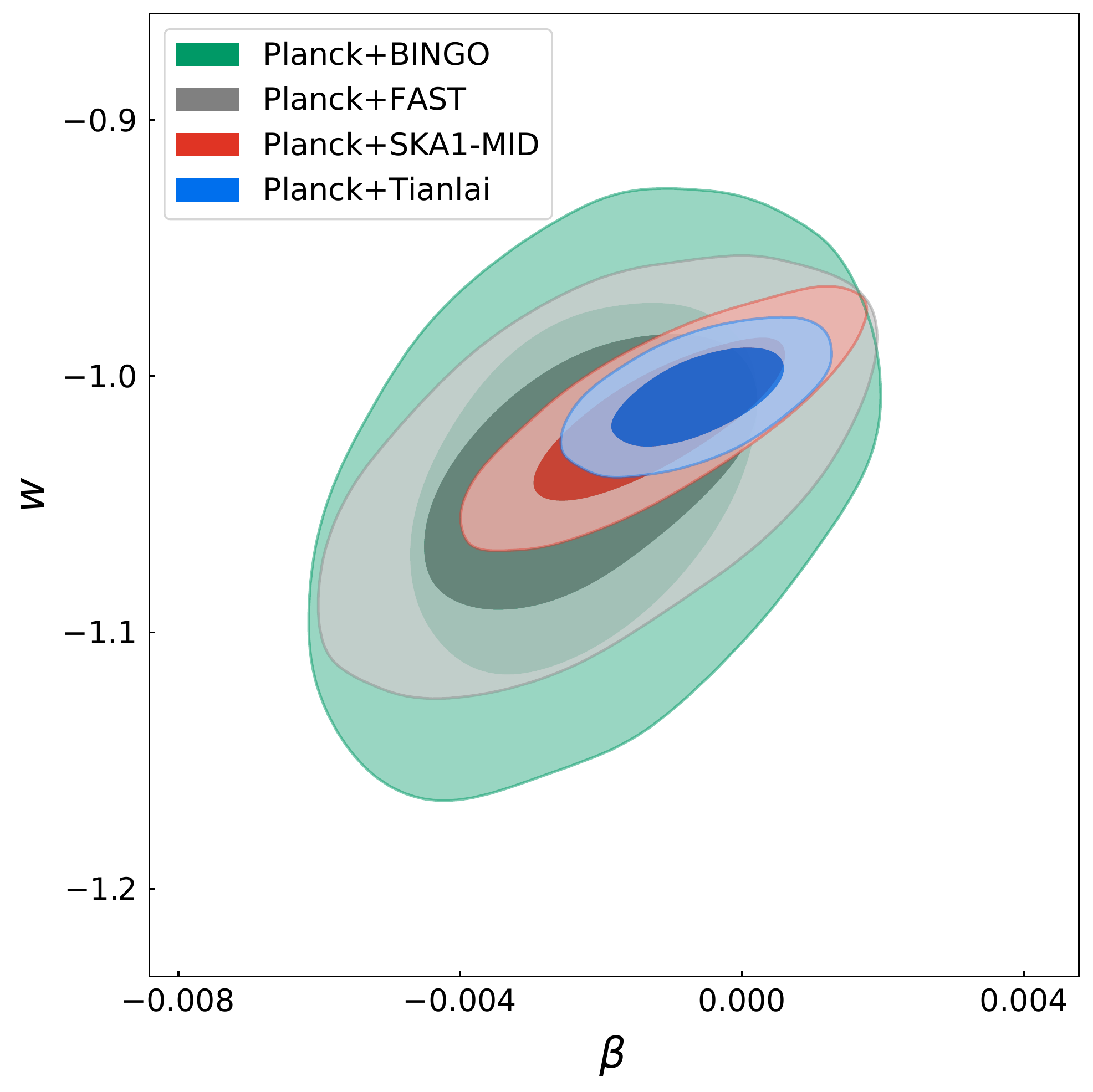}
\caption{Constraints on $\beta$ and $w$ from Planck+BINGO, Planck+FAST, Planck+SKA1-MID, and Planck+Tianlai in the I$w$CDM model.}
\label{IwCDM2}
\end{figure}

In order to compare their constraint abilities in the I$w$CDM model, we show the 1$\sigma$ and 2$\sigma$ measurement error contours for Planck+BINGO, Planck+FAST, Planck+SKA1-MID, and Planck+Tianlai in Figure \ref{IwCDM2}. We obtain $\sigma(\beta)=0.00160$ and $\sigma(w)=0.048$ from Planck+BINGO, which are improved by $8.6\%$ and $84.8\%$, respectively, when combining the BINGO data in the cosmological fit to the Planck CMB data.
This shows that, although BINGO's constraining capability is the weakest among the four 21 cm IM experiments considered in this work, it still can play an important role in breaking the parameter degeneracies in CMB and improving the measurement accuracies of the cosmological parameters (especially $w$) in the I$w$CDM model.
FAST performs slightly better than BINGO, and we obtain $\sigma(\beta)=0.00160$ and $\sigma(w)=0.036$ from Planck+FAST. Evidently, SKA1-MID performs much better than BINGO and FAST, and we obtain $\sigma(\beta)=0.00120$ and $\sigma(w)=0.021$ from Planck+SKA1-MID.
The most stringent constraints on the I$w$CDM model are from Planck+Tianlai, and in this case, we have $\sigma(\beta)=0.00079$ and $\sigma(w)=0.013$. Compared with the Planck result, we find that FAST, SKA1-MID, and Tianlai can improve the constraints on $\beta$ by 8.6\%, 31.4\%, and 54.9\%, respectively.

\section{Conclusion}\label{sec4}

In this work, we investigate the constraint capabilities of the future 21 cm IM experiments for the IDE models. We consider BINGO, FAST, SKA1-MID, and Tianlai as typical examples of 21 cm IM experiments and find that among them, a compact interferometer array like the full-scale Tianlai cylinder array would be the best one in constraining the IDE model.

We find that the 21 cm observations with the full-scale Tianlai cylinder array can tightly constrain $H_0$, $\Omega_{\rm m}$, and $\sigma_8$. For example, in the I$\Lambda$CDM model, the Tianlai-alone data can give the constraint accuracies of $\sigma(H_0)=0.19$ km s$^{-1}$ Mpc$^{-1}$, $\sigma(\Omega_{\rm m})=0.0033$, and $\sigma(\sigma_8)=0.0033$, even much better than those of Planck+optical BAO. 
But, relatively speaking, the Tianlai data cannot provide a tight constraint on the coupling parameter $\beta$ as much as constraints on other cosmological parameters.
However, it is also found that the parameter degeneracy directions from Planck and Tianlai are entirely different; thus, the combination of Planck and Tianlai can well break the parameter degeneracies and give a rather tight constraint on $\beta$. In the I$\Lambda$CDM and I$w$CDM models, we obtain $\sigma(\beta)=2.3\times 10^{-4}$ and $7.9\times 10^{-4}$, respectively, from Planck+Tianlai. This shows that the constraints on $\beta$ can be improved by 90.4\% and 54.9\% in the two models by adding the Tianlai data in the cosmological fit, compared with the case of using only the Planck data.

We also make a detailed comparison for BINGO, FAST, SKA1-MID, and Tianlai in the cosmological-fit study of the IDE models. We find that, for the constraint capability, Tianlai is the best one, and SKA1-MID is slightly less powerful than Tianlai, but both are much better than FAST and BINGO. Our goal is not to show the superiority or inferiority of these experiments against each other but rather to give a global picture of their relative prospects. Our results show that the 21 cm IM experiments will provide a promising tool for exploring the nature of dark energy, and, in particular, a compact interferometer array will play a significant role in measuring the coupling between dark energy and dark matter.

\section*{Acknowledgments}
We thank Xin Wang, Xue-Lei Chen, Ze-Wei Zhao, Ling-Feng Wang, Li-Yang Gao, and Yue Shao for helpful discussions.
This work was supported by the MoST-BRICS Flagship Project (Grant No. 2018YFE0120800), the National Natural Science Foundation of China (Grant Nos. 11975072, 11875102, 11835009, 11690021, 11973047, and 11633004), the National SKA Program of China (Grant No. 2020SKA0110401),
the Chinese Academy of Sciences (CAS) Strategic Priority Research Program (Grant No. XDA15020200),
the Liaoning Revitalization Talents Program (Grant No. XLYC1905011), the Fundamental Research Funds for the Central Universities (Grant No. N2005030), and the Top-Notch Young Talents Program of China (Grant No. W02070050).

\bibliography{Tianlai}
\bibliographystyle{aasjournal}

\end{document}